\documentclass[12pt]{iopart}

\usepackage{graphicx}% Include figure files

\begin{document}

\article{\letter{}}{Critical wetting in power-law wedge geometries}

\author{A Sartori\dag\footnote[1]{To whom correspondence should be addressed
(anna.sartori@curie.fr)} and A O Parry\ddag}

\address{\dag Institut Curie, UMR 168, 11, rue Pierre et Marie Curie, 75005 Paris, France}

\address{\ddag Department of Mathematics, Imperial College, 180, Queen's Gate, London SW7 2BZ}

\ead{anna.sartori@curie.fr}

\ead{a.parry@ic.ac.uk}

\begin{abstract}
We investigate critical wetting transitions for fluids adsorbed in
wedge-like geometries where the substrate height varies as a
power-law, $z(x,y)\sim \vert x \vert ^\gamma$, in one direction.
As $\gamma$ is increased from $0$ to $1$ the substrate shape is
smoothly changed from a planar-wall to a linear wedge. The
continuous wetting and filling transitions pertinent to these
limiting geometries are known to have distinct phase boundaries
and critical singularities. We predict that the intermediate
critical wetting behaviour occurring for $0<\gamma< 1$ falls into
one of {\it{three}} possible regimes depending on the values of
$\gamma,p$ and $q$. The unbinding behaviour is characterised by a
high degree of non-universality, strongly anisotropic correlations
and enhanced interfacial roughness. The shift in phase boundary
and emergence of universal critical behaviour in the linear wedge
limit is discussed in detail.

\end{abstract}


\submitto{\JPCM}

\begin{figure}[h]\hspace{0.5cm}
\includegraphics[width=0.75\textwidth]
{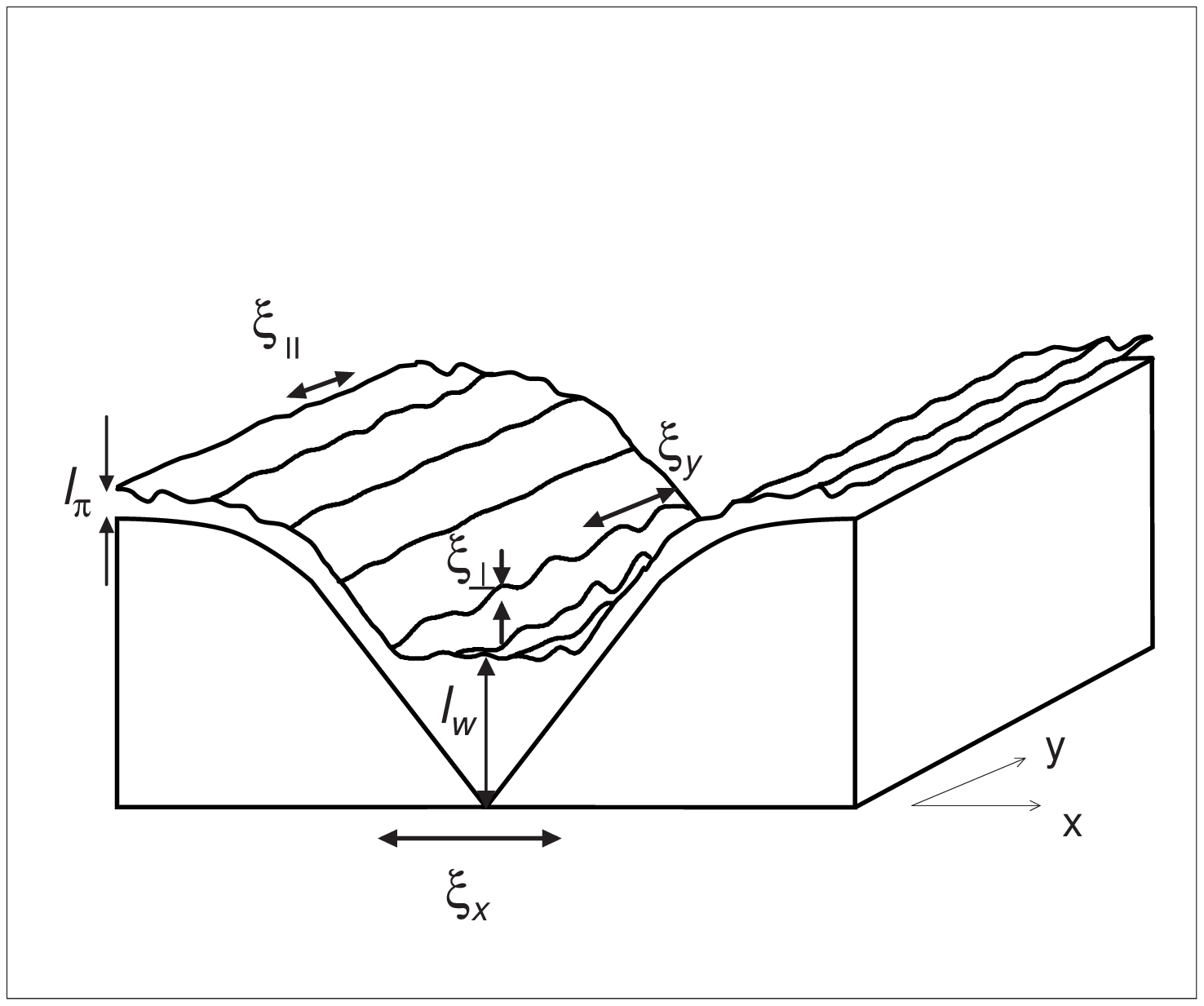} \caption{\label{fig:gammawall} {\small
Schematic illustration of an interfacial configuration in a
generalised wedge geometry. Diverging lengthscales are
highlighted.}}
\end{figure}

\begin{figure}[h]\hspace{0.cm}
\includegraphics[width=0.65\textwidth]
{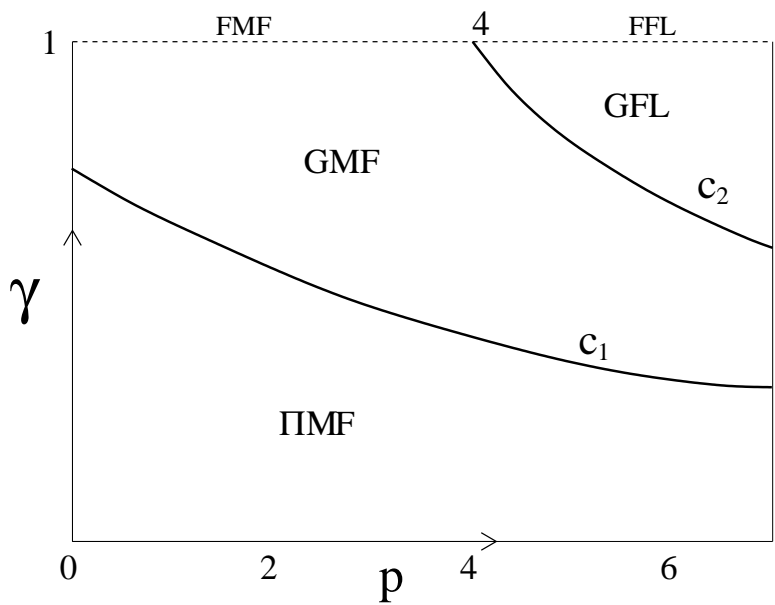} \caption{\label{fig:phdiagr} {\small
Critical regimes for the gamma-wall wedge. The curves $c_1$ and
$c_2$ represent the separatrixes $\gamma_1(q)$ and
$\gamma_2(q,p)$, obtained for $q=p+n$, with $n$ fixed.}}
\end{figure}

\begin{figure}[h]
\includegraphics[width=0.95\textwidth
%,angle=-90
]
{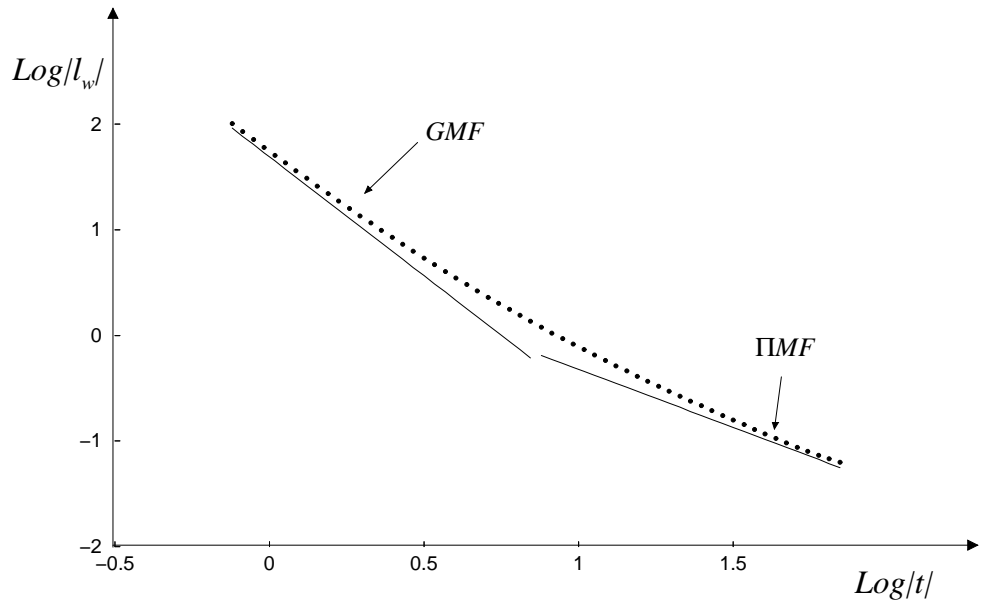}
\caption{\label{5fig:logt-loglo-0.6}{\small Log--Log plot of the
mid-point height $\ell_w$ vs $t$ for $\gamma=3/5$ and dispersion
forces, showing cross--over from planar--like ($\Pi$MF) to GMF
behaviour, with asymptotic criticality at $\beta(3/5)=9/4$.}}
\end{figure}

There is growing interest from theorists and experimentalists in
fluid adsorption on micro-patterned and sculpted solid substrates
\cite{1,2,3,4,5,6,7,8,9,10,11,12,13,14,15}. Surface decoration and
structure can substantially alter the character of fluid
adsorption and lead to novel examples of interfacial phase
transitions and enhanced fluctuation related effects. There is
also strong evidence that there are fundamental connections
between geometry-induced and fluctuation-induced interfacial
phenomena at wedge filling transitions \cite{11} and also between
wedge filling and unzipping transitions for doubled-stranded DNA
\cite{16}. \\
The purpose of the present article is to focus on interfacial
adsorption in generalised 3D wedge shaped geometries, which are
translationally invariant in one direction (along the $y$ axis,
say), but whose height above some reference plane varies as a
power law $z(x,y)\sim \vert x \vert ^\gamma$, for large distances
in the $x$ direction. We refer to this particular class of surface
geometry as the gamma-wall which may be viewed as an example of
deterministic roughness \cite{17,18,19}. Note that by increasing
the exponent $\gamma$ the wall morphology can be changed smoothly
from a planar substrate ($\gamma=0$) to a linear wedge
($\gamma=1$) and eventually to a parallel plate geometry
($\gamma=\infty$). The adsorption properties and interfacial
fluctuation effects in each of these geometries, corresponding to
wetting \cite{20,21}, filling \cite{9,10,11} and capillary
condensation \cite{21}, respectively, are very different to each
other and have received considerable theoretical and experimental
interest. The central question we ask here is, how do the wetting
properties depend on the wall exponent $\gamma$? For substrates
that are completely wet by the fluid (corresponding to vanishing
contact angle $\theta=0$), recent work \cite{12,13} has shown that
the adsorption isotherms for the gamma wall show a sensitive
dependence on $\gamma$, which facilitate the cross-over from
continuous complete wetting ($\gamma=0$) to first-order capillary
condensation ($\gamma=\infty$) phenomena, through a sequence of
novel interfacial behaviours, which emerge at intermediate values
of $\gamma$. Here, we extend this study to the case where the
(planar) substrate undergoes a continuous (critical) wetting
transition at some temperature $T_{\rm w}$ and ask how the phase
boundary and critical exponents characteristic of the interfacial
unbinding depend on $\gamma$. As we shall show this case is
considerably more complex than the complete wetting scenario, due
to the presence of large scale interfacial fluctuation effects,
which gives rise to strongly non-universal critical behaviour and
distinct fluctuation regimes. Nevertheless, the critical
properties characteristic of these regimes are precisely those
that allow us to understand the change in phase boundary and
fluctuation-related properties, that occur as the wall morphology
is changed from a planar substrate
($\gamma=0$) to a linear wedge ($\gamma=1$).\\
{\indent} Consider the interface between a non-planar substrate
modelled as an inert spectator phase, whose height is described by
the a continuous function $z(x,y)$ in the shape of a generalised
wedge (see figure \ref{fig:gammawall}), with the power law
behaviour described above. The substrate is in contact with a bulk
vapour phase at temperature $T$ and pressure $p$, which we will
suppose is tuned to bulk two-phase coexistence $p=p_{\rm sat}(T)$.
The {\it{planar}} wall-fluid interface (corresponding to
$\gamma=0$) is taken to have a continuous wetting transition at
temperature $T_{\rm w}$, at which the contact angle $\theta(T)$
vanishes and the mean interfacial height $l_{\pi}$ diverges. The
mean-field critical singularities occurring at the wetting
transition are found by minimising the binding potential
\begin{equation}
W(l)\,=\,-\frac{t}{l^p}+\frac{b}{l^q}, \label{binding}
\end{equation}
where  $t\propto (T_{\rm w}-T)/T_{\rm w}$, $b>0$ and the exponents
$p$, $q\,
>p$ depend on the ranges of the intermolecular forces. The
critical exponents describing the divergence of the mean
interfacial height $l_{\pi}\sim t^{-\beta_{\rm s}}$, roughness
$\xi_{\perp}\sim t^{-\nu_{\perp}}$ and parallel correlation length
$\xi_{\parallel}\sim t^{-\nu_{\parallel}}$ are given by the well
known expressions
\begin{equation}
\beta_{\rm s}\,=\,\frac{1}{q-p}, \qquad
\nu_{\perp}\,=\,0(\sqrt{\log}),\qquad \nu_{\parallel}\,=\,
\frac{(q+2)}{2(q-p)} \label{wetting}
\end{equation}
and are not altered by thermal fluctuation \cite{20,21}. Note that
the interfacial roughness $\xi_{\perp}$ is negligible compared to
the wetting film thickness. Also recall that the contact angle
vanishes as $\theta\sim t^{(2-\alpha_{\rm s})/2}$, with $2-\alpha_{\rm s}=q/(q-p)$.\\
{\indent} Elementary thermodynamic considerations, based on
balancing bulk and surface tension contributions to the grand
potential, show that the location of the surface phase
boundary/wetting transition in the gamma-wall geometry depends
qualitatively on the behaviour of $r=A_{\gamma}/A_{\pi}$,
corresponding to the ratio of the total to planar (projected) area
of the substrate. Thus, for $0\le\gamma<1$ for which $r=1$, the
wetting phase boundary remains at $T=T_{\rm w}$, whilst for
$\gamma>1$, for which $r=\infty$, the wedge is completely filled
by liquid (gas) at all temperatures such that $\theta(T)$ is less
(greater) than $\pi/2$. Thus, for $\gamma>1$ the wetting
transition at bulk coexistence is superseded by a first-order
capillary condensation-like phenomena. This is closely related to
unbending phase transition occurring on corrugated surfaces
\cite{23}. Hereafter, we restrict our attention to the regime
$\gamma\le 1$ and focus on the fate of the planar wetting
transition as $\gamma$ is
increased from $0$ to $1$. \\
The critical behaviour occurring at the limit $\gamma=1$,
corresponding to the filling of a linear wedge, is known in some
detail \cite{9,10,11}. Writing the wall-function $z(x,y)=\tan
\alpha \,\vert x \vert $, with $\alpha$ the tilt angle, observe
that the ratio $r=\sec \alpha>1$. Accordingly, the phase boundary
for the wedge wetting (filling) is shifted and again thermodynamic
arguments dictate that it occurs at a lower filling temperature
$T_{\rm f}$, satisfying $\theta(T_{\rm f})=\alpha$ rather than
$\theta(T)=0$ \cite{1}. For planar substrates that undergo
critical wetting transitions, the linear wedge filling transition
is also continuous, but is characterised by critical exponents
distinct to those at critical wetting \cite{10,11}. Consider, for
example, the divergence of the wedge mid-point interfacial height
$l_{\rm w}\sim t'^{-\beta_{\rm w}}$, mid-point roughness
$\xi_{\perp}\sim t'^{-\nu_{\perp}}$ and correlation length
$\xi_y\sim t'^{-\nu_y}$, where the latter is measured along the
wedge and $t'\propto (T_{\rm f}-T)/T_{\rm f}$. Calculations based
on effective interfacial models show that the criticality falls
into two regimes \cite{10}. For $p<4$ the critical exponents
belong to a filling mean-field (FMF) regime with critical
exponents $\beta_{\rm w}=1/p$, $\nu_{\perp}= 1/4$,
$\nu_y=1/2+1/p$, in which the roughness is much smaller than the
film thickness, $\xi_{\perp}/l_{\rm w}<<1$. For $p>4$, on the
other hand, there is a filling fluctuation (FFL) regime and the
critical exponents take the universal values
\begin{equation}
\beta_{\rm w}\,=\,1/4,\qquad \nu_{\perp}\,=\,1/4,\qquad
\nu_y\,=\,3/4. \label{fillin2}
\end{equation}
In this regime, $\xi_{\perp}\sim l_{\rm w}$ and the interfacial
fluctuations are controlled by an effective wedge wandering
exponent, so that $l_{\rm w}\sim\xi_{\perp}\sim \xi_y^{\zeta}$ with
$\zeta=1/3$ . The universal value of $\nu_{\perp}$ implies that
fluctuation effects, at the filling of a linear wedge, are always
important and contrast sharply with those at planar
wetting transition.\\
{\indent} With these preliminaries aside, we can now precise the
two central questions addressed in this paper. First, we introduce
critical exponents for the mid-point height, roughness and
correlation length as $t\to 0$ by the identifications
\begin{equation}
l_{\rm w}\sim t^{-\beta(\gamma)},\qquad \xi_{\perp}\sim
t^{-\nu_{\perp}(\gamma)},\qquad \xi_y\sim
t^{-\nu_y(\gamma)}, \label{expdef}
\end{equation}
where we have restricted ourselves to the (unknown) critical
behaviour occurring in the range $0<\gamma<1$. Then, in evaluating
the full $\gamma$, $p$ and $q$ dependence of these critical
exponents, we wish to understand two specific points; {\bf{A}}:
How does the critical behaviour reflect the discontinuous shift of
the phase boundary from $\theta=0$ to $\theta=\alpha$ as
$\gamma\to 1^-$ ? {\bf{B}}: How do the critical exponents change
from their wetting to filling values? In particular, are any
combinations of critical exponents continuous in this limit and
allow us to smoothly `turn on" the fluctuation
effects?\\
{\indent} The starting point for our calculations is the
interfacial Hamiltonian model
 \begin{equation}
H[l] \,=\,\int \int \rmd x\,\rmd y \left[ \frac{\Sigma}{2}
\left(\nabla l \right)^2 +W(l-z(x,y)) \right], \label{effham}
\end{equation}
where $l(x,y)$ is measured relative to the horizontal reference
plane $z=0$, $\Sigma$ denotes the stiffness (surface tension) of
the unbinding liquid-vapour interface and $W(l)$ is the binding
potential (\ref{binding}). The model is only valid for substrate
height functions $z(x,y)$ that have a shallow gradient $\vert
\nabla z \vert <<1$ and thus suffices to determine the critical
behaviour in the regime of interest, $0\le \gamma\le 1$. The
critical exponents are insensitive to the precise nature of the
substrate shape near the $x=0$ line and the power-law shape
may be cut off at some appropriate short-distance. \\
{\indent} The results of our analytical and numerical studies show
that the critical behaviour is strongly non-universal and depends
sensitively on the values of $\gamma,p$ and $q$. We concentrate on
the interpretation of these results, making brief mention of our
calculational details \cite{24} at the end of our article. Figure
\ref{fig:phdiagr} shows how the critical behaviour falls into
three possible categories labelled the planar mean-field
($\Pi$MF), geometrical mean-field (GMF) and geometrical
fluctuation (GFL) regimes, respectively. Within the $\Pi$MF
regime, $0\le\gamma<\gamma_1(q)$ with
\begin{equation}
\gamma_1(q)\,=\,\frac{2}{2+q}, \label{gam1}
\end{equation}
the substrate shape does not alter the values of the wetting
critical exponents and $\beta(\gamma)$, $\nu_{\perp}(\gamma)$ and
$\nu_y(\gamma)$ are identical to $\beta_{\rm s}$, $\nu_{\perp}$ and
$\nu_{\parallel}$ shown in (\ref{wetting}). For $\gamma>\gamma_1$,
corresponding to the geometrical, GMF and GFL regimes, on the
other hand, the wedge geometry alters the critical exponents from
their planar values. In both these regimes the interfacial height
diverges with a modified exponent
\begin{equation}
\beta(\gamma)\,=\,\frac{(2-\alpha_{\rm s})}{2}\,\,\frac{\gamma}{(1-\gamma)}.
\label{betagam}
\end{equation}
{\indent} Note that the critical exponent $\beta(\gamma)>\beta_{\rm s}$,
for $\gamma>\gamma_1$, so that in the limit $t \to 0$ the
mid-point height $l_{\rm w} \gg l_{\pi}$. In figure
\ref{5fig:logt-loglo-0.6} we show numerical results for the
mid--point height for the case of non--retarded van der Waals
forces ($p=2$, $q=3$) and $\gamma=3/5$, for which we predict
$\beta(3/5)=9/4$. Notice that the initial divergence of the
interfacial height is planar-like (with critical exponent
$\beta_{\rm s}=1$), but crosses over to the asymptotic geometrical
result as $t\to 0$. The boundary between the GMF and $\Pi$MF
regimes occurs when $\beta(\gamma_1)=\beta_{\rm s}$, so there is smooth
cross--over in the critical behaviour for the interfacial height
at the $\Pi$MF/GMF separatrix. The GMF and GFL regimes are
distinguished from each other by the behaviour of the correlation
length critical exponents $\nu_{\perp}(\gamma)$ and
$\nu_y(\gamma)$. For $p<4$ the GMF regime extends from $\gamma_1$
to the linear wedge limit $\gamma=1^-$. For $p>4$, on the other
hand, the GMF regime terminates at $\gamma=\gamma_2$, where
\begin{equation}
\gamma_2(q,p)\,=\,\frac{2p}{q(p-4)+2p}. \label{gam2}
\end{equation}
Within the GMF regime, the roughness is still small compared to
the film thickness, but is larger than in the $\Pi$MF regime and
$\nu^{\rm GMF}_{\perp}(\gamma)>0$. There is pronounced
non-universality in this regime with
\begin{equation}
\nu^{\rm GMF}_{\perp}(\gamma)\,=\,\frac{1}{4}+\frac{\beta(\gamma)}
{4}\left[p-\frac{2(1-\gamma)}{\gamma}\right] \label{nuperp}
\end{equation}
and
\begin{equation}
\nu^{\rm GMF}_y(\gamma)
\,=\,\frac{1}{2}+\beta(\gamma)\left(1+\frac{p}{2}\right).
\label{nuy}
\end{equation}
There is a smooth change from the $\Pi$MF to the GMF regime for
the fluctuation critical exponents so that, at the separatrix,
$\nu^{\rm GMF}_{\perp}(\gamma_1)=0$. For $\gamma>\gamma_2$
(relevant for systems with $p>4$ only), corresponding to the GFL
regime, on the other hand, the wedge wetting transition in the
gamma wall is fluctuation dominated and we can identify
\begin{equation}
\nu_{\perp}^{\rm GFL}\,=\,\beta(\gamma),\qquad
\nu_{\perp}^{\rm GFL} \,=\,\zeta(\gamma)\cdot \nu^{\rm GFL}_y,
\label{GFL}
\end{equation}
where $\zeta(\gamma)=\gamma/(\gamma+2)$ is the wedge wandering
exponent for the gamma wall. Thus, in the GFL regime one has
simple scaling relations between the diverging lengthscales,
$l_{\rm w}\sim\xi_{\perp}\sim \xi_y^{\zeta({\gamma})}$. For systems with
purely short-ranged forces, the GFL regime spans the entire range
$0<\gamma<1$ and $\beta(\gamma)=
\nu_{\parallel}\,\gamma/(1-\gamma)$. \\
Returning to the more general case observe again there is smooth
cross-over between the behaviour of the fluctuation critical
exponents at the separatrix between the GMF and GFL regimes with,
for example, $\nu_{\perp}^{\rm GMF}(\gamma_2)=\beta(\gamma_2)$.
The geometrical regimes are also characterised by a strong degree
of anisotropy, with the correlation length $\xi_y$ much greater
than the lateral extent of the filled region $\xi_x \, \sim\,
l_{\rm w}^{1/\gamma}$. Interfacial fluctuations within the GMF and GFL
regimes are pseudo-one dimensional, in contrast with $\Pi$MF regime. \\
{\indent} At this point, a number of remarks are in order.\\
{\indent}(I) The existence of three fluctuation regimes for
critical wetting in the generalised wedge geometry contrasts with
the case of complete wetting \cite{12,13}, for which there are
only two and no significant enhancement of fluctuation effects.
Intriguingly, the number of regimes for critical and complete
wetting in the gamma-wall wedge are the same as that induced by
thermal (or impurity induced) fluctuation effects at planar
critical and complete wetting transitions, respectively \cite{21}.\\
{\indent} (II) In the geometry affected regimes (GMF and GFL), the
equilibrium profile $l_{\rm eq}(x)$ has a particularly simple
structure which can be seen to directly lead to the critical
exponent identification (\ref{betagam}). Near the center of the
wedge, the interface is flat and at near constant height $l_{\rm
eq}(x)\sim l_{\rm w}$. At some distance $\xi_x\sim l_{\rm w}^{1/\gamma}$, the
interface strikes the wall and thereafter closely follows its
shape. The interfacial height at the wedge mid-point is determined
by the simple condition that the local angle of incidence between
the interface and wall is equal to the contact angle. Observe
that, as $\gamma \to 1^-$, the critical exponent for the
interfacial height diverges. Including amplitude factors, we find
that the mid-point height diverges as $l_{\rm w} \sim
(t_{\rm w}/t)^{\beta(\gamma)}$, with $t_{\rm w}$ a non-universal constant. As
$\gamma \to 1^-$ this implies that the height becomes macroscopic
for all $t<t_{\rm w}$, which represents a shift of the phase boundary
from $t=0$ to $t=t_{\rm w}$. This is equivalent to the shift of the
phase boundary from $\theta=0$ to $\theta=\alpha$ for linear
wedge filling and answers our first question {\bf{A}}.\\
{\indent} (III) Some aspects of the interfacial fluctuations show
a smooth change from wetting to filling-like behaviour, as
$\gamma$ is increased, and allow us to give a quantitative answer
to question {\bf{B}}. This is most simply seen in the wedge
wandering exponent $\zeta(\gamma)=\gamma/(\gamma+2)$ pertinent to
the GFL regime, which generalises the linear wedge result
$\zeta=1/3$. Less obvious is the behaviour of the critical
exponent ratio $\beta(\gamma)/\nu_{\perp}(\gamma)$, which also
recovers the linear wedge result, such that
\begin{equation}
\lim_{\gamma\to 1^-}
\frac{\beta(\gamma)}{\nu_{\perp}(\gamma)}\,=\,\frac{\beta_{\rm w}}{\nu_{\perp}},
\label{lim}
\end{equation}
where the RHS is equal to $4/p$ and $1$ for $p<4$ and $p>4$,
respectively.\\
{\indent} For systems with non-retarded van der Waals forces ({\it
i.e.} with binding potential exponents $p=2$, $q=3$) we make the
following predictions. The planar result pertinent to the standard
critical wetting transition $\beta_{\rm s}=1$, $\nu_{\perp}=0$ and
$\nu_{\parallel}=5/2$ are unchanged within a $\Pi$MF regime
corresponding to $0<\gamma <2/5$. For $1>\gamma>2/5$ the
transition belongs to the GMF and the critical exponents are
geometry sensitive. For example, at $\gamma=1/2$ we predict
\begin{equation}
\beta(1/2)\,=\,\frac{3}{2},\qquad
\nu_{\perp}(1/2)\,=\,\frac{1}{4},\qquad
\nu_y(1/2)\,=\,\frac{7}{2}.
\label{nonret}
\end{equation}
{\indent} The value of the roughness critical exponent
$\nu_{\perp}(1/2)=1/4$ is significant, since it is
{\it{independent}} of the value of $q$ and is therefore also valid
for tricritical wetting. This degree of universality is similar to
the true universality of $\nu_{\perp}$ predicted for linear wedge
filling.\\
{\indent} To finish, we make brief mention of the methods used in
our calculations. In both the $\Pi$MF and GMF regimes the
roughness is much smaller than the interfacial height and
mean-field methods are appropriate. Numerical results obtained by
minimising (\ref{effham}) are complemented by analytical
approaches following approximate solution to the Euler-Lagrange
equation, based on standard variational methods. This is
straightforward in the GMF regime, since the profile has a
particularly simple structure. The results for the correlation
length critical exponent $\nu_y$ and roughness exponent
$\nu_{\perp}$ were first obtained by solving the Ornstein-Zernike
equation for the structure factor $S(Q)$, corresponding to the
Fourier transform of the mid-point height-height correlation
function $\langle l(y_1)l(y_2)\rangle$ with respect to
wave-vectors along the wedge. The critical behaviour in the GMF
and GFL regimes can also be described using an effective
one-dimensional model Hamiltonian $H_{\gamma}[l]$, which describes
the energy-cost of constrained interfacial configurations in terms
of the mid-point height $l(y)=l(x=0,y)$ only. The model can be
derived from the underlying interfacial model (\ref{effham}) using
standard methods, which have been previously applied to the linear
wedge problem \cite{10}. The dimensional reduction, explicit in
this method, is justified by the extreme anisotropy of
fluctuations in the GMF
and GFL regimes as $t \to 0$.   \\
The reduced dimensional effective interfacial Hamiltonian has the
form
\begin{equation}
H_{\gamma}[l]\,=\,\int \rmd y \left[ \sigma \,l^{1/\gamma}
\left(\frac {\rmd y }{\rmd l} \right)^2 + V_{\gamma}(l) \right],
\label{effeone}
\end{equation}
where $\sigma$ is a non-universal constant (proportional to
$\Sigma$) and $V_{\gamma}(l)$ is the wedge binding potential.
Minimization of $V_{\gamma}(l)$ identically recovers the
mean-field expression for $l_{\rm w}$ in the $\Pi$MF and GMF regimes.
For large $l$ this has the expansion
\begin{equation}
V_{\gamma}(l)\propto l^{1/\gamma} \left[
\left(\theta^2-c_{\gamma} \,l^{2(1-\gamma)/\gamma}
\right)+d_{\gamma}\,l^{-p} \right], \label{last}
\end{equation}
where $c_{\gamma}$ and $d_{\gamma}$ are non-universal constants.
In the limit $\gamma \to 1^-$, we find $c_{\gamma}\to \alpha^2$
and $d_{\gamma} \to t/(p-1)$, so that both $V_{\gamma}(l)$ and
(\ref{effeone}) smoothly recover the linear wedge model considered
in \cite{10}. From the one-dimensional model it is straightforward
to derive all the critical exponents quoted earlier using standard
methods.\\
{\indent} In summary, we have investigated the geometry dependence
of critical wetting exponents for fluids adsorbed in power-law
wedges. Our results show that surface shape has both a stronger
(and subtler) effect on critical wetting than complete wetting
transitions,
 with criticality falling into three possible regimes
which facilitate the crossover from planar wetting
to linear wedge filling.\\

\ack The authors are grateful to Dr Carlos Rasc\`{o}n and Dr
A.J.Wood for very helpful discussions. A.S. wishes to thank the
EPSRC for financial support.

\newpage

\vspace{1cm}

\end{document}